\documentclass[prb]{revtex4}
\usepackage{amssymb}
\usepackage{graphicx}
\usepackage{amsfonts}
\usepackage{amsmath}
\usepackage{color}

\begin{document}
\title{Response functions in multicomponent Luttinger liquids}
\author{E. Orignac}
\affiliation{Laboratoire de Physique de l'ENS-Lyon, CNRS UMR 5672, 46
  All\'ee d'Italie, 69364 Lyon Cedex 07, France}
\author{R. Citro}
\affiliation{Dipartimento di Fisica "E.R. Caianiello", Universit\`a
  degli Studi di Salerno, and Spin-CNR UO of Salerno, Via Ponte Don Melillo 84084 Fisciano (Sa)
  Italy}
\date{\today}

\begin{abstract}
We derive an analytic expression for the zero temperature Fourier
transform of the density-density correlation function of a
multicomponent Luttinger liquid with different velocities. By
employing Schwinger identity and a generalized Feynman identity exact
integral expressions are derived, and approximate analytical forms are
given for frequencies close to each component singularity. We find
power-like singularities and compute the corresponding
exponents. Numerical results are shown for $N=3$  components and
implications for experiments on cold atoms are discussed.
\end{abstract}
\maketitle

\section{Introduction}
\label{sec:intro}

It is well known that as a result of spin-charge separation,
interacting one-dimensional spin-1/2 fermions with repulsive
interaction at incommensurate filling form a two-component Luttinger
liquid (LL).  Multicomponent LL with more than two
components can be obtained in fermionic systems with repulsive
interaction\cite{kawaguchi2000,assaraf_su(n),miyashita2002}.  Possible
realization of such systems are provided by multichannel quantum
wires\cite{yoshioka2011}, carbon
nanotubes\cite{egger_nanotube_theory,egger_nanotubes} and biased bilayer graphene\cite{killi2010}. Interaction of
acoustic phonons with spin-1/2 fermions in one dimension can also give
rise to a multicomponent LL.\cite{loss1994,martin1995}
In Mott insulating materials, a multicomponent LL can be formed in spin-orbital
chains\cite{khomskii_orbital_degeneracy,azaria_su4,li_su4,li_su4_long,yamashita_su4, yamashita_orbital_dmrg, lee_spin_orbital_mf, frischmuth_su4,gu_su4} and spin tubes
\cite{degier2000} under the effect
of an applied magnetic field\cite{orignac_spintube,yu_spin_orbital}
and in spin-1 chains with biquadratic
interactions\cite{fath_biquadratic,manmana2010}.
 More recently, atom trapping
technology has permitted the realization of Bose-Fermi
mixtures,\cite{best2009,guenter2006,ferlaino2006,klempt2007} as well
as degenerate gases with internal degrees of
freedom.\cite{desalvo2010,taie2010} In the latter case, it has been
suggested theoretically that these systems could realize SU(N) spin systems\cite{cazalilla2009,hermele2009,gorshkov2010,xu2010} in low dimensions.
In parallel, techniques for trapping atoms in one dimension have been
developed\cite{bouchoule2009,kinoshita_tonks_continuous,paredes_tonks_optical,stoferle_tonks_optical,moritz_oscillations,koehl_nofk,palzer2009,haller2010}.
In Bose-Fermi mixtures trapped in
one dimension, a multicomponent LL
behavior is
expected\cite{mathey2004,sengupta2005,luscher2009,rizzi2008,mathey07_bose_fermi,mathey07_mixture,mathey2007,takeuchi2007,mathey2007a,pollet2006,hu2006,takeuchi2006,pollet05_bosefermi,takeuchi2005,frahm2005}. Similarly,
multicomponent systems with repulsive interactions are also expected
to exhibit in one dimension a multicomponent LL behavior\cite{ulbricht2010,he2011}
The real space correlation function of the multicomponent Luttinger
liquid can be readily
obtained\cite{muttalib1986,loss1994,mathey2007a}.  However, the
majority of experimental observables are actually Fourier transform
(FT) of these correlation functions and thus the FT of the
multicomponent LL has to be obtained.  Because of the branch cut
structure of the correlation functions in a LL, this is a non-trivial
task. In the single component case, the calculation can be done in
closed
form\cite{luther_bosonisation,schulz_correlations_1d,schu_spins}. For
two component systems, only the exponents of the power law
singularities could be
predicted.\cite{voit_spectral,meden_spectral,nakamura_suzumura}
Recently a closed form of the $2k_F$ component of the density-density
response function in a two-component LL (i.e.  the spin-1/2 case with
different charge and spin velocites) at zero temperature was
obtained\cite{iucci2007} in terms of Appell hypergeometric
functions.\cite{erdelyi_functions_1,dlmf_appell} Such an expression
permits the description of the crossovers between the different power
law singularities of the response functions.  In the present
manuscript, we derive an exact expression for the FT of
the density-density correlation function
in the general case of a
multicomponent LL with different velocities for the modes.
We show
that in this general case, the FT of the Matsubara
correlation functions are expressed in terms of
the Srivastava-Daoust generalized hypergeometric
functions. We give  give the full analytical
continuation of the correlation functions
to real frequencies, recovering the leading power-law
singularities and  describing the various crossovers between them.

  The paper is so
organized. In Sec. \ref{sec:model} we give the Hamiltonian of a
general multicomponent system and derive the general expression for
the Matsubara correlation functions at zero temperature.
In the subsections \ref{sec:schwinger}
and~\ref{sec:feynman} we derive two exact integral representations  of the Fourier transforms of the Matsubara correlation functions by means of a Schwinger identity and a Feynman
identity, respectively. The integral representation obtained from the Schwinger identity is used to predict the exponents of the power-law singularities.
In Sec.\ref{sec:analytic-feynman}, starting from the integral representation derived from the Feynman identity,  we obtain
the analytic continuation of the Matsubara correlator in various cases
and give the asymptotic expression close to the singular
points. Finally, we give some conclusions in Sec.\ref{sec:concl}.

\section{Model} \label{sec:model}
In the present paper,
we wish to consider the case of a general multicomponent system.
The continuum Hamiltonian is:
\begin{eqnarray}
  \label{eq:bare-ham}
  H&=&\sum_{a=1}^N \int dx \left[ -\psi^\dagger_a
    \frac{\hbar^2}{2M_a} \partial_x^2 \psi_a\right] \nonumber \\
   &&+ \sum_{1\le a<b \le N} \int dx dx' V_{ab}(x-x') \rho_a(x)
   \rho_a(x'),
\end{eqnarray}
where $\psi_a$ annihilates a particle of type $a$,
$\rho_a(x)=\psi^\dagger_a \psi_a$  is the corresponding particle
density, and $M_a$ the corresponding mass. The interaction of
particles of type $a$ with particles of type $b$ is $V_{ab}$.
The particles may be either bosons or fermions. The
Hamiltonian~(\ref{eq:bare-ham}) can be
bosonized\cite{orignac2010_mix} and its expression reads:
\begin{eqnarray}
  \label{eq:hamiltonian}
  H=\sum_{a,b} \int \frac{dx}{2\pi} \left[ (\pi \Pi_a) M_{ab} (\pi
    \Pi_b) + (\partial_x \phi_a) N_{ab} (\partial_x \phi_b) \right],
\end{eqnarray}
where $[\phi_a(x),\Pi_b(x')]=i\delta_{ab} \delta(x-x')$,
and the matrices $M$ and $N$ can be obtained respectively from
the variation of the ground state energy with change of boundary
conditions and particle numbers.\cite{orignac2010_mix}
The Hamiltonian~(\ref{eq:hamiltonian}) can be
diagonalized, and the equal time correlations
can then be obtained.\cite{orignac2010_mix}
We wish to calculate the Matsubara time dependent correlation
functions. In the case of density correlations, since the density is
expressed as:
\begin{eqnarray}
  \label{eq:density}
  \rho_a(x) = -\frac 1 \pi \partial_x \phi_a +
  \sum_{m=-\infty}^{+\infty} A_m \cos [2m (\phi_a(x) - \pi  \rho_a^{0} x)],
\end{eqnarray}
We will need to calculate correlation functions of the form $\langle
T_\tau e^{i m \phi_a(x,\tau)} e^{-i m \phi_a(0,0)} \rangle$, where $T_\tau$ is the Matsubara time ordered operator. If we
have bosonic particles, the bosonized form of the annihilation
operator being:
\begin{eqnarray}
  \label{eq:annihilator}
  \psi_a(x)= e^{i \theta_a(x)} \left[\sum_m B_m e^{i 2m [\phi_a(x) - \pi
      \rho_a^{0} x]} \right],
\end{eqnarray}
where $\nabla \theta_a =\pi \Pi_a$,
 the leading term in the single-particle Green's function is
 proportional to
$\langle T_\tau e^{i \theta_a (x,\tau)} e^{-i \theta_a (x,\tau)}
\rangle$. Because of the duality transformation\cite{orignac2010_mix}
 $M \leftrightarrow N$ and $\theta_a \leftrightarrow \phi_a$, it is
 sufficient to calculate the correlation functions of the form $\langle T_\tau e^{i \sum_n \alpha_n \phi_n(x,\tau)}  e^{-i \sum_n
    \alpha_n \phi_n(0,0)}\rangle$ and use the duality transformation
  to obtain the correlation functions $\langle T_\tau e^{i \sum_n \alpha_n \theta_n(x,\tau)}  e^{-i \sum_n
    \alpha_n \theta_n(0,0)}\rangle$.
We have:
\begin{eqnarray}
\langle T_\tau e^{i \sum_n \alpha_n \phi_n(x,\tau)}  e^{-i \sum_n
    \alpha_n \phi_n(0,0)}\rangle = \exp \left[ -\sum_{n<m} \alpha_n
    \alpha_m G_{nm}(x,\tau) \right],
\end{eqnarray}
with\cite{orignac2010_mix},
\begin{eqnarray}
  \label{eq:green}
  G_{nm}(x,\tau)=\pi \sum_{\omega_n} \int \frac{dq}{2\pi} e^{-|q|
    \alpha}
  (1-e^{i(qx-\omega_n \tau)}) [(\omega_n^2 + (MN) q^2)^{-1} M]_{nm}.
\end{eqnarray}
Let us introduce the projection operator $P_\lambda$ that projects on
the eigenspace of eigenvalue $u_\lambda^2$ of the matrix $MN$ to
rewrite the matrix $(\omega_n^2 + (MN) q^2)^{-1}$ as:
\begin{eqnarray}
  (\omega_n^2 + (MN) q^2)^{-1} = \sum_{n}
  \frac{P_n}{\omega_n^2 + u_n^2 q^2}
\end{eqnarray}
In the limit of $\beta \to \infty$, we have:
\begin{eqnarray}
  \int \frac{d\omega}{2\pi} (\omega^2 + (MN) q^2)^{-1} (1-e^{i(qx
    -\omega \tau)})  e^{-|q|
    \alpha}  =  \sum_\lambda  \frac{P_\lambda}{2
    u_\lambda |q|} (1-e^{i qx -u_\lambda |q \tau|})  e^{-|q|
    \alpha},
\end{eqnarray}
So that:
\begin{eqnarray}
  G(x,\tau)= \frac 1 4 \sum_\lambda \frac{P_\lambda}{u_\lambda} M \ln \left(\frac{x^2 +
      (u_\lambda |\tau|+\alpha)^2}{\alpha^2}\right),
\end{eqnarray}
and thus at zero temperature:
\begin{eqnarray}
  \label{eq:corr-func}
  \langle T_\tau e^{i \sum_n \alpha_n \phi_n(x,\tau)}  e^{-i \sum_n
    \alpha_n \phi_n(0,0)}\rangle = \prod_\lambda \left(\frac{\alpha^2}{x^2 +
      (u_\lambda |\tau| +\alpha)^2}\right)^{\eta_\lambda},
\end{eqnarray}
with:
\begin{eqnarray}\label{eq:eta-exponent}
  \eta_\lambda = \sum_{a,b} \alpha_a \frac{(P_\lambda M)_{ab}}{u_\lambda} \alpha_b.
\end{eqnarray}
In the case of the $\theta$ correlation function, one obtains a
formula analogous to (\ref{eq:corr-func}), with the exponents $\eta_n$
replaced by $\bar{\eta}_n$ where:
\begin{eqnarray}\label{eq:etabar}
  \bar{\eta}_\lambda = \sum_{a,b} \alpha_a \frac{(Q_\lambda N)_{ab}}{u_\lambda}
  \alpha_b,
\end{eqnarray}
$Q_\lambda$ being the projector on the eigenstate of $NM$ having the
eigenvalue $u_\lambda^2$. We note that for $\tau=0$, using
Eqs.~(\ref{eq:corr-func})--(\ref{eq:eta-exponent}), we recover the
expression of the exponents of equal-time correlations from
\cite{orignac2010_mix}. These formulas are particularly useful when implementing numerical methods like DMRG and exact diagonalization as one could use the following results to predict the response function from
ground state energy computations.
Knowing the Matsubara time ordered Green's function,
Eq.~(\ref{eq:corr-func}),  we wish to obtain the corresponding Matsubara response
function:
\begin{eqnarray}
  \label{eq:matsubara}
 \chi^M(q,i\omega_n)=\int dx d\tau e^{i (qx -\omega_n \tau)}  \langle T_\tau e^{i \sum_n \alpha_n \phi_n(x,\tau)}  e^{-i \sum_n
    \alpha_n \phi_n(0,0)}\rangle,
\end{eqnarray}
and the retardated response function
$\chi(q,\omega)=\chi^M(q,i\omega_n\to \omega+i0)$. In the following,
we  use two complementary
approaches, the first one based on the Schwinger identity\cite{lebellac_qft} in
Sec.~\ref{sec:schwinger}, that will allow us to predict the
singularities of the retardated response function,
and the second one based on the Feynman
identity\cite{lebellac_qft} that will allow us to make a connection
with the results for the two-component case\cite{iucci2007}.
We will assume that we can take the limit
$\alpha \to 0$ in the integrals, i. e. $\eta=\sum \eta_n <1$.

Let us remind that for the density-density correlation function its FT is related to the scattering cross section $\sigma$ of light at a frequency $\omega$ and angle $\Omega$ incident on a sample, by the relation
\begin{equation}
\frac{d^2 \sigma}{d\omega d\Omega}\propto S(q,\omega)=\mathrm{Sign}(\omega) \mathfrak{Im} \chi (q,\omega)
\end{equation}
where $S(q,\omega)$ is the dynamic structure factor. This quantity is
accessible e.g. by means of inelastic neutron/light scattering when
spin/density fluctuations are induced in the system and their subsequent
relaxation is measured.

\subsection{Schwinger identity}
\label{sec:schwinger}
The Schwinger identity is\cite{lebellac_qft}:
\begin{eqnarray}
  \frac{1}{(x^2+(u \tau)^2)^\alpha}=\int_0^{+\infty}
  \frac{d\lambda}{\Gamma(\alpha)} \lambda^{\alpha-1} e^{-\lambda
    (x^2+(u \tau)^2)}
\end{eqnarray}

In the multicomponent case, it allows us to rewrite the
expression~(\ref{eq:corr-func}) as:
\begin{eqnarray}
  \prod_n \left(\frac{\alpha^2}{x^2 +
      (u_n |\tau| +\alpha)^2}\right)^{\eta_n} = \int \prod_n
  \frac{d\lambda_n \lambda_n^{\eta_n-1}}{\Gamma(\eta_n)}
  \exp\left[-x^2 \left(\sum_n \lambda_n\right) - \tau^2 \left(\sum_n
      \lambda_n u_n^2 \right)  \right]
\end{eqnarray}

The Fourier transformation in (\ref{eq:matsubara}) reduces to a
Gaussian integral, and we find:
\begin{eqnarray}\label{eq:schwinger-rep}
  \chi(q,i\omega_n)=\pi \alpha^{2\sum \eta_n} \int  \prod_n
  \frac{d\lambda_n
    \lambda_n^{\eta_n-1}}{\Gamma(\eta_n)}\frac{e^{-\frac 1 4 \left[
        \frac{q^2}{ \left(\sum_n \lambda_n\right)} + \frac{\omega_n^2}{\left(\sum_n
      \lambda_n u_n^2 \right)} \right]}}{\sqrt{\left(\sum_n
    \lambda_n\right) \left(\sum_n
      \lambda_n u_n^2 \right)}}
\end{eqnarray}

With the change of variables:
\begin{eqnarray}
  \lambda_j=\lambda_1 \mu_{j-1} (j=2,\ldots,N),
\end{eqnarray}
we rewrite (\ref{eq:schwinger-rep}) as:
\begin{eqnarray}
  \label{eq:schwinger-interm1}
  \chi(q,i\omega_n)&=&\pi \alpha^{2\sum \eta_n} \int
  \frac{d\lambda_1}{\Gamma(\eta_1)} \lambda_1^{\sum \eta_n -2}
  \nonumber \\
& & \int \prod_{j=1}^{N-1} \frac{d\mu_j
  \mu_j^{\eta_{j+1}-1}}{\Gamma(\eta_{j+1})} \frac{e^{-\frac 1
    {4\lambda_1} \left[\frac{q^2}{1+\sum_{j=1}^{N-1} \mu_j} + \frac
      {\omega^2}{u_1^2 +\sum_{j=1}^{N-1} \mu_j u_{j+1}^2}
    \right]}}{\sqrt{\left(1+\sum_{j=1}^{N-1} \mu_j \right) \left(u_1^2
        +\sum_{j=1}^{N-1}
        \mu_j u_{j+1}^2 \right) }}
\end{eqnarray}
With the change of variable $\lambda_1=\mu^{-1}$, the integration over
$\lambda_1$ can be done in closed form. We find, provided that $\sum
\eta_n<1$:
\begin{eqnarray}
  \label{eq:schwinger-final}
  \chi(q,i\omega_n)&=&\pi \alpha^{2\sum \eta_n} \frac{\Gamma(1-\sum_j
    \eta_j)}{\prod_j \Gamma(\eta_j)} \int \prod_{j=1}^{N-1} d\mu_j \mu_j^{\eta_{j+1}-1}\frac{\left[\frac{q^2}{1+\sum_{j=1}^{N-1} \mu_j} + \frac
      {\omega^2}{u_1^2 +\sum_{j=1}^{N-1} \mu_j u_{j+1}^2}
    \right]^{\sum \eta_n -1}} {\sqrt{\left(1+\sum_{j=1}^{N-1} \mu_j \right) \left(u_1^2
        +\sum_{j=1}^{N-1}
        \mu_j u_{j+1}^2 \right) }}
\end{eqnarray}

We can perform the analytic continuation on (\ref{eq:schwinger-rep})
by substituting $i\omega_n \rightarrow \omega+i0$. When $\omega \simeq u_1 q$, we
have to consider the integral:
\begin{eqnarray}
  \label{eq:scwhinger-asymp}
   \int \prod_{j=1}^{N-1} d\mu_j \mu_j^{\eta_{j+1}-1} \left[ u_1^2 q^2
     -\omega^2
       +\sum_{j=1}^{N-1} \mu_j (u_{j+1}^2 q^2 -
      {\omega^2})
    \right]^{\sum \eta_n -1}
\end{eqnarray}
With the change of variables $\mu_j =(u_1^2 q^2 -\omega^2)\xi_j $,
we finally find that for $|\omega| \to u_1 q$,
\begin{eqnarray}
\label{eq:limitchi}
  \chi(q,\omega) \sim | (u_1 q)^2 -\omega^2|^{2(\sum \eta_n) -\eta_1
    -1},
\end{eqnarray}
provided $2(\sum \eta_n) -\eta_1 -1<0$. In the general case, we expect
to find $\chi(q,\omega\simeq u_j q ) \sim \sim | (u_j q)^2
-\omega^2|^{2(\sum_n \eta_n) -\eta_j -1}$ when  $2(\sum \eta_n) -\eta_j
-1<0$.
We note that $\eta=\sum_n \eta_n$ is the exponent of the equal time
correlation functions.\cite{orignac2010_mix} The exponents of the
singularities satisfy a ``sum rule'':
\begin{eqnarray}
  \sum_{n=1}^N (2 \eta - \eta_n -1) = (2 N -1) \eta - N
\end{eqnarray}
\subsection{Feynman identity}
\label{sec:feynman}

Using the identity from \cite{lebellac_qft} (Appendix B), we can
rewrite the correlation function~(\ref{eq:corr-func}) as a multiple
integral:

\begin{eqnarray}
  \label{eq:feynman}
  \prod_{j=1}^N \frac 1 {(x^2+u_n^2 \tau^2)^{\eta_n}} =
  \frac{\Gamma(\sum_1^2 \eta_n)}{\prod_{j=1}^N \Gamma(\eta_j)} \int
  \prod_{j=1}^N  dw_j w_j^{\eta_j-1} \delta(1-\sum_{j=1}^N w_j)
  \left(x^2 +\sum_{j=1}^N w_j u_j^2 \tau^2\right)^{-\left(\sum_{j=1}^N \eta_j\right)}
\end{eqnarray}

The Matsubara response
function~(\ref{eq:matsubara}) is then found from
the integral(11.4.16) of Ref.~\onlinecite{abramowitz_math_functions}:
\begin{eqnarray}
& &  \int dx d\tau e^{i (qx -\omega_n \tau)} \left(x^2 +\sum_{j=1}^N w_j
  u_j^2
    \tau^2\right)^{-\eta} \nonumber \\
&=& \frac{2^{2(1-\eta)} \pi \Gamma(1-\eta)}{\Gamma(\eta) \sqrt{\sum_{j=1}^N w_j
  u_j^2}}  \left(q^2 + \frac{\omega_n^2}{\sum_{j=1}^N w_j
  u_j^2}\right)^{\eta-1},
\end{eqnarray}
where $1/4<\eta<1$ as:
\begin{eqnarray}
  \label{eq:feynman-interm1}
  \chi(q,i\omega_n)=\frac{\Gamma(1-\eta)}{\prod_{j=1}^N
    \Gamma(\eta_j)} \int \prod_{j=1}^N dw_j w_j^{\eta_j-1}
  \delta(1-\sum_{j=1}^N w_j)  \frac{2^{2(1-\eta)} \alpha^{2\eta} \pi}{ \left(\sum_{j=1}^N w_j
  u_j^2\right)^{\eta-1/2} }  \left( \omega_n^2 + \sum_{j=1}^N w_j
  u_j^2 q^2  \right)^{\eta-1},
\end{eqnarray}
where we have defined $\eta=\sum_j \eta_j$. For the case of
$\eta=1/2$, the integral~(\ref{eq:feynman-interm1}) can be expressed
as a Lauricella function $F_D$, which actually reduces to a simple
product. The result is simply:
\begin{eqnarray}
  \label{eq:special-case}
  \chi(q,i\omega_n) = 2\pi \alpha \prod_{j=1}^N (\omega^2 + u_j^2
  q^2)^{-\eta_j}.
\end{eqnarray}
This result generalizes the one obtained for the two-component case in
Ref.~\onlinecite{iucci2007}.

When $\eta>1/2$, we can use again the Feynman
identity\cite{lebellac_qft} to write:
\begin{eqnarray}
  \frac 1 {\left(\frac{\omega^2}{q^2} + \sum_j u_j^2 w_j\right)^{1-\eta}
      \left(\sum_j u_j^2 w_j\right)^{\eta-1/2}} =
      \frac{\Gamma(1/2)}{\Gamma(\eta-1/2)\Gamma(1-\eta)} \int_0^1 ds
    s^{-\eta}(1-s)^{\eta-3/2} \left( s \frac{\omega^2}{q^2} + \sum_j
      u_j^2 w_j\right)^{-1/2},
\end{eqnarray}
and find:
\begin{eqnarray}
  \label{eq:feynman-interm2}
  \chi^{(M)}(q,i\omega_n)=\frac{\pi 2^{2(1-\eta)}\alpha^2  \Gamma(1/2)}{\Gamma(\eta-1/2)
    \Gamma(\eta)} (q\alpha)^{2(\eta-1)} \int_0^1 ds && \frac{s^{-\eta}
    (1-s)^{\eta-3/2}}{\left(\frac{\omega^2}{q^2}s + u_N^2\right)^{1/2}} \times \nonumber \\
  && F_D^{(N-1)}\left(\frac 1 2 ; \eta_1,\ldots,
    \eta_{N-1};\eta;\frac{u_N^2-u_1^2}{u_N^2 + s
      \frac{\omega^2}{q^2}}, \ldots, \frac{u_N^2-u_{N-1}^2}{u_N^2 + s
      \frac{\omega^2}{q^2}}\right),
\end{eqnarray}
where the label $(M)$ stands for the Matsubara correlation function and $F_D^{(N-1)}$ is a Lauricella hypergeometric function of $N-1$
variables. The integral~(\ref{eq:feynman-interm2}) can be expressed in
closed form\cite{srivastava1995}  using Srivastava-Daoust generalized
hypergeometric series\cite{exton1976}
 of $N$ variables as:
\begin{eqnarray}
  \label{eq:srivastava-daoust}
   \chi^{(M)}(q,i\omega_n)=F(\eta) (q\alpha)^{2(\eta-1)}
  F_{1;0,\ldots,0,1}^{1;1,\ldots,1,1}\left[\begin{array}{rrrrr}(1/2;1,\ldots,1,1):&
      \eta_1,&\ldots&,\eta_{N-1},& 1-\eta \\ (\eta;1,\ldots,1,0):&
      -,&\ldots,&-,&1\end{array};1-\frac{u_1^2}{u_N^2},\ldots,1-\frac{u_{N-1}^2}{u_N^2},-\frac{\omega^2}{u_N^2
    q^2} \right],\nonumber \\
    \end{eqnarray}
where $F(\eta)=\frac{\pi 2^{2(1-\eta)}\alpha^2  \Gamma(1-\eta))}{\Gamma(\eta) u_N}$. When $N=2$, the Srivastava-Daoust hypergeometric series reduces to
an Appell $F_2$ hypergeometric function of 2 variables. Using the
identity (16.16.3) from [\onlinecite{dlmf_appell}], this function is seen to
reduce to the  $F_1$ Appell function expression given in
Ref.~[\onlinecite{iucci2007}].
Since we are not aware of any study of the  analytic
continuation of the Srivastava-Daoust hypergeometric series
outside their circle of convergence, we will not pursue with
Eq.~(\ref{eq:srivastava-daoust}). Instead, we will consider directly the
analytic continuation of the integral (\ref{eq:feynman-interm2}).
Introducing the new variable:
\begin{eqnarray}
  \label{eq:feynman-chvar}
  t=\frac{\omega_n^2+u_N^2 q^2}{s\omega_n^2+u_N^2 q^2}s,
\end{eqnarray}
the integral (\ref{eq:feynman-interm2}) is rewritten:
\begin{eqnarray}
  \label{eq:feynnman-final}
   && \chi^{(M)}(q,i\omega_n)=\frac{\pi 2^{2(1-\eta)}\alpha^2  \Gamma(1/2)}{\Gamma(\eta-1/2)
    \Gamma(\eta) u_N} \left(\frac{\omega_n^2\alpha^2} {u_N^2}+
    (q\alpha)^2 \right)^{(\eta-1)} \int_0^1 dt t^{-\eta}
    (1-t)^{\eta-3/2} \times   \nonumber \\
    &&
  F_D^{(N-1)}\left(\frac 1 2 ; \eta_1,\ldots,
    \eta_{N-1};\eta;\frac{u_N^2-u_1^2}{u_N^2}\left(1-
      \frac{\omega_n^2 t}{\omega_n^2 +u_N^2q^2}\right), \ldots,;\frac{u_N^2-u_{N-1}^2}{u_N^2}\left(1-
      \frac{\omega_n^2 t}{\omega_n^2 + u_N^2q^2}\right) \right),
\end{eqnarray}
In the case $N=2$, the Lauricella function in the
integral~(\ref{eq:feynnman-final}) reduced to a ${}_2F_1$ Gauss
hypergeometric function. Doing the integral gives a $F_1$ Appell
hypergeometric function\cite{erdelyi_functions_1}.
With $N=3$, the Lauricella
function reduces to a $F_1$ Appell function.

In the case of $\eta<1/2$, the integral (\ref{eq:feynman-interm2}) is
divergent. In order to obtain a convergent integrals, we write in
Eq.~(\ref{eq:feynman-interm1}):
\begin{eqnarray}
  \frac{\left(\omega^2 + q^2 \sum_j u_j^2 w_j\right)^{\eta-1}}{
    \left(\sum_j u_j^2 w_j\right)^{\eta-1/2}} = \sum_{j=1}^N
  \frac{u_j^2 w_j}{\left(\omega^2 + q^2 \sum_j u_j^2
      w_j\right)^{1-\eta} \left(\sum_j u_j^2 w_j\right)^{\eta+1/2}},
\end{eqnarray}
and apply the Feynman identity to each term in the sum.
We thus find:
\begin{eqnarray}
  \label{eq:chi-eta-gt-half}
  && \chi(q,i\omega_n)= \pi \alpha^{2\eta} q^{2\eta-2} 2^{2(1-\eta)}
    \frac{\Gamma(3/2)}{\Gamma(\eta+1/2)\Gamma(\eta+1)}\nonumber \\ &&
    \times  \int_0^1 ds
    \frac{s^{-\eta}(1-s)^{\eta-1/2}}{\left(s \frac{\omega^2}{q^2} +
        u_N^2\right)^{3/2}} \left[ \sum_{\ell=1}^N \eta_\ell u_\ell^2
      F_D^{(N-1)} \left(\frac 3 2; \{\eta_j +\delta_{j\ell}\}_{1\le
          j\le
          N-1};\eta+1;\frac{u_N^2-u_1^2}{s\frac{\omega^2}{q^2}+u_N^2},\ldots,
       \frac{u_N^2-u_1^2}{s\frac{\omega^2}{q^2}+u_N^2}\right) \right]
\end{eqnarray}
Each term in the sum is then expressible with Srivastava-Daoust
hypergeometric functions.

\subsection{Analytic continuation of the Matsubara correlator}
\label{sec:analytic-feynman}

To obtain the retardated response function, we have to find the
analytic continuation  $i\omega_n \rightarrow \omega + i\epsilon$
 of (\ref{eq:feynnman-final}). We will first discuss the special case
 of $\eta=1/2$, where the continuation is straightforward,
leading to  a simple picture of the behavior of the retardated response
 function. Then, we will turn to the more complicated case of
 $\eta>1/2$, for which the calculations are more involved. We will see
 however that the simple picture of the case $\eta=1/2$ is preserved.

\subsubsection{The case of $\eta=1/2$}
\label{sec:eta-1-2}

In the case of $\eta=1/2$, the analytic
 continuation is easily obtained from Eq.~(\ref{eq:special-case}).
We have for $u_j q < \omega < u_{j+1} q$:
\begin{eqnarray}
  \label{eq:spec-case-contin}
  \chi(q,\omega+i0) = 2\pi \alpha e^{i\pi \sum_{\ell=1}^j \eta_\ell}
  \prod_{j=1}^N |\omega^2 -(u_j q)^2|^{-\eta_j}
\end{eqnarray}

\begin{figure}
  \centering
 \includegraphics[width=9cm]{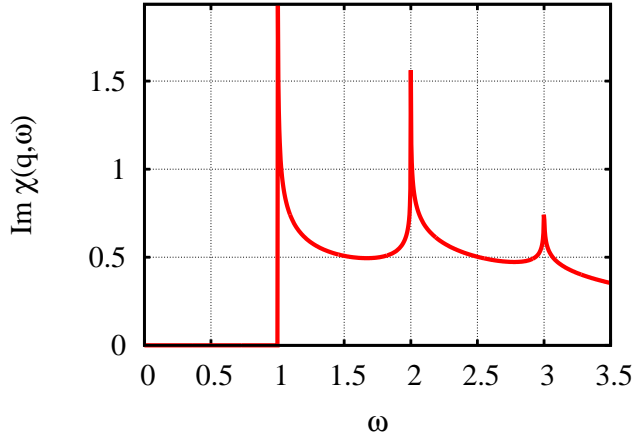}
  \caption{The imaginary part of the response function
    $\chi(q,\omega)$ for the case $\eta_1=0.25$,$\eta_2=0.15$ and
    $\eta_3=0.1$. The velocities are $u_2=2u_1$ and $u_3=3u_1$. The
    unit of frequency $\omega$ is $u_1q$. The unit of  $\mathfrak{Im}
    \chi(q,\omega)$ is $2\pi\alpha/(u_1 q)$. Power-law divergences are obtained
    for $\omega=u_{1,2,3}q$.}
  \label{fig:eta-half}
\end{figure}

So that:
\begin{eqnarray}
  \mathfrak{Im} \chi(q,\omega \to u_j q + 0) \sim 2 \pi \alpha \sin
  \left[\pi\sum_{l=1}^j \eta_l\right]  |\omega^2 -(u_j q)^2|^{-\eta_j}
  \prod_{l\ne j} |(u_j^2 -u_l^2)
  q^2 |^{-\eta_l},
\end{eqnarray}
and:
\begin{eqnarray}
    \mathfrak{Im} \chi(q,\omega \to u_{j+1} q - 0) \sim 2 \pi \alpha \sin
  \left[\pi\sum_{l=1}^j \eta_l\right] |\omega^2 -(u_{j+1}
  q)^2|^{-\eta_{j+1}}  \prod_{l\ne j+1} |(u_{j+1}^2 -u_l^2)
  q^2 |^{-\eta_l},
\end{eqnarray}
showing that the spectral function has singularities with exponent
$-\eta_j$ everytime $\omega \sim u_j q$. This result is is agreement
with the result of Sec.~\ref{sec:schwinger} for
$\eta=1/2$. The spectral function has a threshold for $\omega< u_1 q$
as there are no excitations of the system having a lower energy.
Moreover, we note that for $j>1$:
\begin{eqnarray}
\frac{\mathfrak{Im} \chi(q,\omega \to u_j q + 0)}{\mathfrak{Im}
  \chi(q,\omega \to u_j q - 0)} = \frac{ \sin
  \left[\pi\sum_{l=1}^j \eta_l\right]} { \sin
  \left[\pi\sum_{l=1}^{j-1} \eta_l\right]},
\end{eqnarray}
which implies that the peaks of the spectral functions are asymmetric
around $\omega = u_j q$. The imaginary part of the response function
is represented on Fig.~\ref{fig:eta-half} for the case $\eta=1/2$.

\subsubsection{The case of $\eta>1/2$}
\label{sec:eta-gt-1-2}

In the general case, we need to find the analytic continuation of
Eq.~(\ref{eq:feynnman-final}) for $\eta>1/2$ or
(\ref{eq:chi-eta-gt-half}) for $\eta<1/2$. Since the method is similar
in the two cases, we will concentrate on the case of $\eta>1/2$.

Formally, under the analytic continuation, the variables in
Eq.~(\ref{eq:feynnman-final}) become:
\begin{eqnarray}
  \label{eq:contin-variable}
  \frac{u_N^2-u_j^2}{u_N^2}\left(1-\frac{\omega_n^2 t}{\omega_n^2 +
      u_N^2 q^2} \right) \to
  \left(1-\frac{u_j^2}{u_N^2}\right)\left(1+\frac{(\omega+i\epsilon)^2}{(u_N q)^2
      -(\omega+i\epsilon)^2} t\right),
\end{eqnarray}

The Lauricella function $F_D$ has cuts every time the real part of one
of the variables is larger than one.
In the case of $\omega > u_N q$, the real part of all the variables,
acconding to (\ref{eq:contin-variable}) will remain less than one,
and the analytic continuation is straightforward. We can then derive
an equivalent for the Lauricella function in the limit of $\omega \to
u_N q + 0$. We find:
\begin{eqnarray}
 &&  F_D^{(N-1)}\left(\frac 1 2 ; \eta_1,\ldots,
    \eta_{N-1};\eta;\frac{u_N^2-u_1^2}{u_N^2}\left(1-
      \frac{\omega^2 t}{\omega^2 -u_N^2q^2}\right), \ldots,;\frac{u_N^2-u_{N-1}^2}{u_N^2}\left(1-
      \frac{\omega^2 t}{\omega^2 - u_N^2q^2}\right) \right) \nonumber
  \\
&& \sim
  \frac{\Gamma(\eta) \Gamma(\eta_N+1/2 -\eta)}{\Gamma(1/2)
    \Gamma(\eta_N)} \left(\frac{\omega^2 -(u_N q)^2}{\omega^2
      t}\right)^{\eta - \eta_N} \prod_{j=1}^{N-1}
  \left(1-\frac{u_J^2}{u_N^2}\right)^{-\eta_j},
\end{eqnarray}
yielding:
\begin{eqnarray}
  \label{eq:omega-gt-un-q}
  \mathfrak{ Im} \chi(q,\omega) = \frac{\pi 2^{1-\eta} \alpha^2
    \Gamma(\eta_{N+1}-2\eta)\sin (\pi \eta)}{\Gamma(\eta_N) u_N}
  \frac{(\omega^2 - (u_N q)^2)^{2\eta - \eta_N
      -1}}{(\omega^2)^{\eta-\eta_N}
    \left(\frac{u_N^2}{\alpha^2}\right)^{\eta-1}} \prod_{j=1}^{N-1}
  \left(1-\frac{u_j^2}{u^2_N}\right)^{-\eta_j}.
\end{eqnarray}
In Eq.~(\ref{eq:omega-gt-un-q}), the exponent predicted in
Sec.~\ref{sec:schwinger} is recovered. For $\omega \to u_N q -0$, it
is also possible to find an asymptotic estimation of the Lauricella
function in the form:
\begin{eqnarray}
&&  F_D^{(N-1)}\left(\frac 1 2 ; \eta_1,\ldots,
    \eta_{N-1};\eta;\frac{u_N^2-u_1^2}{u_N^2}\left(1-
      \frac{\omega^2 t}{\omega^2 -u_N^2q^2}\right), \ldots,;\frac{u_N^2-u_{N-1}^2}{u_N^2}\left(1-
      \frac{\omega^2 t}{\omega^2 - u_N^2q^2}\right) \right) \nonumber
  \\
&& \sim
  \frac{\Gamma(\eta) \Gamma(\eta_N+1/2 -\eta)}{\Gamma(1/2)
    \Gamma(\eta_N)} \left(\frac{(u_N q)^2 -\omega^2 }{\omega^2
      t}\right)^{\eta - \eta_N} e^{i \pi (\eta -\eta_N)\mathrm{sign}(\omega)} \prod_{j=1}^{N-1}
  \left(1-\frac{u_J^2}{u_N^2}\right)^{-\eta_j},
\end{eqnarray}
giving:
\begin{eqnarray}
  \label{eq:omega-alt-un-q}
  \mathfrak{ Im} \chi(q,\omega \to u_N q-0) \sim \frac{\pi 2^{1-\eta} \alpha^2
    \Gamma(\eta_{N+1}-2\eta)\sin [\pi (\eta-\eta_N)]}{\Gamma(\eta_N) u_N}
  \frac{( (u_N q)^2 - \omega^2 )^{2\eta - \eta_N
      -1}}{(\omega^2)^{\eta-\eta_N}
    \left(\frac{u_N^2}{\alpha^2}\right)^{\eta-1}} \prod_{j=1}^{N-1}
  \left(1-\frac{u_j^2}{u^2_N}\right)^{-\eta_j},
\end{eqnarray}
i. e. the same power law divergence as in Eq.~(\ref{eq:omega-gt-un-q})
but with a different prefactor. The ratio of the two expressions is:
$\sin[\pi(\eta-\eta_N)]/\sin(\pi \eta)$, as previously noted in the
special case of $\eta=1/2$.
For $\epsilon \to 0$, the imaginary part of (\ref{eq:contin-variable})
is positive for $\omega < u_N q$
and the real part of (\ref{eq:contin-variable}) is equal to one
for $t=t_j=\frac{(u_N
  q/\omega)^2-1}{(u_N/u_j)^2-1}$. We have $t_1<t_2<\ldots<t_{N-1}$.

 In
particular,  one finds that when $\omega < u_1 q$, $t_1>1$,
so that no analytic continuation of the Lauricella function under the
integral sign  in (\ref{eq:feynnman-final}) is
needed. The response function remains purely real in that case, and
the spectral function vanishes.
In the case $u_j q < \omega < u_{j+1} q$, we find that $t_1<\ldots <
t_j<1<t_{j+1}<\ldots<t_{N-1}$. As a result, the integral has to be
split into a sum of integrals over the intervals $[0,t_1]$,
$[t_l,t_{l+1}]$ with $1\le l \le j-1$ and $[t_j,1]$. For each
interval, the analytic continuation of the Lauricella function
Eq.~(\ref{eq:lauricella_continuation})  must be
used in order to express the full integral.  To give a concrete
example of the procedure, we will at first focus on
the case $N=3$. We need to consider the following integral:

\begin{eqnarray}
  \label{eq:feynnman-part}
{\text I}(q,\omega)= \int_0^1 dt t^{-\eta}
    (1-t)^{\eta-3/2}
  F_1\left(\frac 1 2 ; \eta_1,
    \eta_{2};\eta;\frac{u_3^2-u_1^2}{u_3^2}\left(1-
      \frac{\omega_n^2 t}{\omega_n^2 +u_3^2q^2}\right);\frac{u_3^2-u_{2}^2}{u_3^2}\left(1-
      \frac{\omega_n^2 t}{\omega_n^2 + u_3^2q^2}\right) \right).
\end{eqnarray}

For $\omega<u_1 q$, we have:

\begin{eqnarray}
{\text I}(q,\omega)= \int_0^1 dt t^{-\eta}
    (1-t)^{\eta-3/2}
  F_1\left(\frac 1 2 ; \eta_1,
    \eta_{2};\eta;\frac{u_3^2-u_1^2}{u_3^2}\left(1+
      \frac{\omega^2 t}{ u_3^2q^2-\omega^2}\right);\frac{u_3^2-u_{2}^2}{u_3^2}\left(1+
      \frac{\omega^2 t}{u_3^2q^2-\omega^2}\right) \right).
\end{eqnarray}

Let us now consider the case of $u_1 q < \omega < u_2 q$.
First, we split the integral with the rule given above and consider
the intervals $t\in \lbrack 0, \frac{\frac{(u_3
    q)^2}{\omega^2}-1}{\frac{u_3^2}{u_1^2}-1}\rbrack$ and $t\in
\lbrack \frac{\frac{(u_3
    q)^2}{\omega^2}-1}{\frac{u_3^2}{u_{1}^2}-1},1\rbrack$. In the
first interval, the analytic continuation is straightforward. In the
second integral, we must use (\ref{eq:x2_less_1}).

 To
calculate the integrals, it is convenient to perform the following change of variables in the first and second interval respectively:
\begin{eqnarray}
t&=&\frac{\frac{u_3 q}{\omega}-1}{\frac{u_3}{u_1}-1} s_1\nonumber \\
t&=& \frac{\frac{u_3 q}{\omega}-1}{\frac{u_3}{u_1}-1}+ \left( 1-\frac{\frac{u_3 q}{\omega}-1}{\frac{u_3}{u_1}-1}\right)s_2 \nonumber
\end{eqnarray}
In the integration over $s_1$, we don't need an analytic continuation of the
Appell function. In the $s_2$ integration, one of the two variables is
larger than $1$ and we find the analytic continuation using
Eq.~(\ref{eq:x2_less_1}).

 Since we are interested in calculating the imaginary part of
 (\ref{eq:feynnman-part}), only the term proportional to  $e^{i\pi \eta_1}$ gives a contribution and we can perform explicitly the integral when $|\omega|\rightarrow u_1q$. The final results for the imaginary part of the integral (\ref{eq:feynnman-part}) yields:
\begin{eqnarray}\label{eq:u1q-lt-w-lt-u2q}
&& \mathfrak{Im }{\text I}(\omega,q)=\frac{\pi \Gamma(\eta)}{\Gamma(1/2)
  \Gamma(1/2+\eta-\eta_1)}
\frac{\left(\frac{u_1^2}{u_3^2}\right)^{\eta-1/2-\eta_1}}{\left(\frac{u_3^2-u_1^2}{u_3^2}\right)^{\eta-\eta_2-1}
  \left(\frac{u_2^2-u_1^2}{u_3^2}\right)^{\eta_2}}
\left(\frac{\frac{\omega^2}{u_1^2}-q^2}{q^2 -
    \frac{\omega^2}{u_3^2}}\right)^{2\eta-1-\eta_1}
\left(\frac{\frac{u_3^2}{u_1^2}-1}{\frac{u_3^2
      q^2}{\omega^2}-1}\right)^{1/2}
\nonumber \\ && \times \int_0^1
\frac{ds s^{\eta-\eta_1-1/2} (1-s)^{\eta-3/2}}{ \left( 1+\frac{\frac{\omega^2}{u_1^2}-q^2}{q^2 -
    \frac{\omega^2}{u_3^2}}s \right)^\eta \left(1+\frac{\omega^2-u_1^2
    q^2}{u_3^2 q^2 -\omega^2}s\right)^{\eta-1} } F_1\left(1-\eta_1;\frac 1
2,\eta_2;\eta+\frac 1 2-\eta_1; - \frac{\omega^2-u_1^2
    q^2}{u_3^2 q^2 -\omega^2}s; \frac{u_3^2-u_2^2}{u_2^2-u_1^2}\frac{\omega^2-u_1^2
    q^2}{u_3^2 q^2 -\omega^2}s  \right)
\end{eqnarray}
When $\omega \to u_1 q$, the integral is behaving as $(\omega^2
-u_1^2 q^2)^{2\eta-1-\eta_1}$, in agreement with (\ref{eq:limitchi}).
For $\omega \to u_2 q$, we need to consider the behavior of the Appell
function as one of its arguments is going to unity, while the other is
negative. Using the results from App.~\ref{app:asymp-f1}, we find that
when $\eta_1<1/2$, we can use Eq.~(\ref{eq:asymp-f1-power}) to
approximate Eq.~(\ref{eq:u1q-lt-w-lt-u2q}) as:
\begin{eqnarray}
  \label{eq:w-to-u2q-minus}
 \mathfrak{Im} I(\omega,q) &\propto& \int_0^1 ds s^{\eta-3/2}
 \left(\frac{u_3^2-u_1^2}{u_2^2-u_1^2}\frac{(u_2q)^2-\omega^2}{(u_3q)^2-\omega^2}
   -s\right)^{\eta-\eta_2-1/2} \nonumber \\
&\propto&
\left(\frac{u_3^2-u_1^2}{u_2^2-u_1^2}\frac{(u_2q)^2-\omega^2}{(u_3q)^2-\omega^2} \right)^{2\eta-\eta_2-1}
\end{eqnarray}
Again, this result is in agreement with the power law divergence
expected from (\ref{eq:limitchi}).

We now turn to the case $u_2 q < \omega < u_3 q$. In that case, we
have to split the integral in Eq.~(\ref{eq:feynnman-part}) into three
integrations. The first one, on
$[0,\frac{(u_3q/\omega)^2-1}{(u_3/u_1)^2-1}]$ does not contribute to
the imaginary part. The second one, on the interval
$[\frac{(u_3q/\omega)^2-1}{(u_3/u_1)^2-1},\frac{(u_3q/\omega)^2-1}{(u_3/u_2)^2-1}]$,
requires an analytic continuation of the $F_1$ function using
(\ref{eq:x2_less_1}) and gives a contribution to the imaginary
part:
\begin{eqnarray}
 &&  \frac{\pi \Gamma(\eta) (1-(u_1/u_3)^2)^{1/2} }{\Gamma(1/2) \Gamma(\eta_1)
    \Gamma(1/2+\eta-\eta_1)} \left(\frac{u_3^2}{u_1^2}\right)^{2-\eta}
  \left(\frac{u_2^2-u_1^2}{u_3^2-u_2^2}\right)^{\eta-\eta_1+1/2}
  \left(\frac{u_3^2-u_1^2}{u_2^2-u_1^2}\right)^{\eta_2}
   \left(\frac{(u_3 q)^2 -\omega^2}{\omega^2-(u_1
      q)^2}\right)^{\eta-1} \left(1-\frac{u_1^2
      q^2}{\omega^2}\right)^{-1/2} \nonumber \\
&& \times \int_0^1 ds s^{\eta-\eta_1-1/2}
\left(1+\frac{u_2^2-u_1^2}{u_3^2-u_2^2} s\right)^{1-\eta}
\left(1+\frac{u_3^2}{u_1^2} \frac{u_2^2 -u_1^2}{u_3^2-u_2^2}
  s\right)^{-\eta} \left(1-\frac{u_2^2-u_1^2}{u_3^2-u_2^2} \frac{(u_3 q)^2 -\omega^2}{\omega^2-(u_1
      q)^2} s\right)^{\eta-3/2} \nonumber \\ && \times  F_1\left(1-\eta_1;\frac 1 2,
    \eta_2;\eta+\frac 1 2 -\eta_1; -\frac{u_2^2-u_1^2}{u_3^2-u_2^2}
    s,s\right)
\end{eqnarray}

 The last integration, on
$[\frac{(u_3q/\omega)^2-1}{(u_3/u_2)^2-1},1]$, requires an analytic
continuation of $F_1$ using (\ref{eq:x_1-x_2-gt1}), and contributes
two terms.
The first one is:
\begin{eqnarray}
&&  \frac{\pi \Gamma(\eta) \Gamma(1-\eta_2) |\omega|
    (u_3^2)^{2\eta-\eta_1-1} (u_2^2)^{-\eta} (u_1^2)^{\eta_1+1/2-\eta}}{\Gamma(1/2) \Gamma(\eta_1)
    \Gamma(\eta-1/2) \Gamma(1-\eta_1-\eta_2)}
  \frac{(u_2^2-u_1^2)^{\eta+1/2-2\eta_1-\eta_2}}{(u_3^2-u_2^2)^{1-\eta_1-\eta_2}
    (u_3^2-u_1^2)^{\eta-\eta_1-1}} \frac{(\omega^2 - (u_2
    q)^2)^{\eta-1/2}}{((u_3 q)^2 -\omega^2)^\eta} \nonumber \\
&&\times \int_0^1 ds (1-s)^{\eta-3/2} \left(1+\frac{u^2_3}{u^2_2}
  \frac{\omega^2 -(u_2 q)^2}{(u_3 q)^2 -\omega^2} s\right)^{-\eta}
\left(1+\frac{\omega^2 -(u_2 q)^2}{(u_3 q)^2 -\omega^2} s
  \right)^{1-\eta}  \left(1+\frac{u^2_3-u_1^2}{u^2_2-u_1^2}
  \frac{\omega^2 -(u_2 q)^2}{(u_3 q)^2 -\omega^2}
  s\right)^{\eta-\eta_1-1/2} \nonumber \\ && F_1\left(1-\eta_1;\frac 1 2,\frac 3 2
  -\eta;2-\eta_1-\eta_2; \frac{u^2_1-u_2^2}{u^2_3-u_2^2}, \frac{1}{1+\frac{u^2_3-u_1^2}{u^2_2-u_1^2}
  \frac{\omega^2 -(u_2 q)^2}{(u_3 q)^2 -\omega^2}
  s}\right),
\end{eqnarray}
and for $\eta-\eta_2-1/2<0$ behaves as $(\omega^2 -(u_2
q)^2)^{2\eta-\eta_2-1}$ in agreement with~(\ref{eq:limitchi}).
The second one is:
\begin{eqnarray}
  && \frac{\Gamma(\eta)\Gamma(1-\eta_2) \sin [ \pi (\eta_1+\eta_2)] }{
    \Gamma(1/2) \Gamma(1/2+\eta-\eta_2)}
  \left(\frac{u_3^2-u_2^2}{u_2^2-u_1^2}\right)^{\eta_1} \left(
    \frac{\omega^2}{u_2^2} \frac{u_3^2-u_2^2}{(u_3
      q)^2-\omega^2}\right)^\eta   \left(\frac{u_3^2}{\omega^2} \frac{\omega^2-(u_2
      q)^2} {u_3^2-u_2^2}\right)^{\eta-1/2} \left(\frac{\omega^2-(u_2
      q)^2}{(u_3
      q)^2-\omega^2} \right)^{\eta-\eta_2-1/2}
\nonumber \\
&& \times \int_0^1 s^{\eta-\eta_2-1/2}(1-s)^{\eta-3/2}
\left(1+\frac{u_3^2}{u_2^2}\frac{\omega^2-(u_2
      q)^2}{(u_3
      q)^2-\omega^2}s\right)^{-\eta}\left(1+\frac{\omega^2-(u_2
      q)^2}{(u_3
      q)^2-\omega^2}s\right)^{-1/2}\nonumber \\ && \times F_1\left(1-\eta_2;\frac 1
    2,\eta_1;\eta+\frac 1 2 -\eta_2;-\frac{\omega^2-(u_2
      q)^2}{(u_3
      q)^2-\omega^2} s, - \frac{u_3^2-u_1^2}{u_2^2-u_1^2} \frac{\omega^2-(u_2
      q)^2}{(u_3
      q)^2-\omega^2} s\right)
\end{eqnarray}
The latter term contributes a divergence
$(\omega^2-(u_2q)^2)^{2\eta-\eta_2-1}$ as $\omega\to u_2 q+0$  in
agreement with (\ref{eq:limitchi}).

To summarize, the qualitative behavior of the spectral function is the
same as in the special case of $\eta=1/2$. The spectral function has a
threshold at $\omega=u_1q$, and has power law singularities for
$\omega=u_j q$ with an exponent given by Eq.~(\ref{eq:limitchi}).

%\subsection{analysis of singularities}
%\label{sec:singular-feynman}

%=====================================================================================fig2
\begin{figure}[t]
\vspace{1.0cm}
\centering
\includegraphics[scale=0.7]{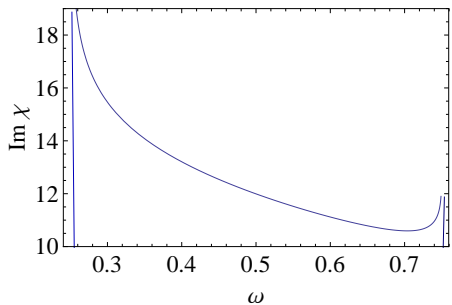}
\caption{Spectral function close to the singularity for N=2 and the values of the parameters: $\eta_1=0.3,\eta_2=0.1,\eta=0.6,u_1=0.25,u_2=0.75$.}
\label{fig:fig2}
\end{figure}
%=========================================================================================

The case of  a general $N$ is  treated in the
Appendix~\ref{app:general-n}. It is found that the leading singularity
for $\omega \to u_j q$ is again a power law divergence, with exponent
given by (\ref{eq:limitchi}).

\section{Conclusion}
\label{sec:concl}

We derived a analytical expressions for the
zero temperature Fourier
transform of the density-density correlation function and the bosonic Green's function of a
multicomponent Luttinger liquid with different velocities. By using
both a Schwinger identity and a generalized Feynman identity, we derived
 exact integral representations
while an approximate analytical form was
given for frequencies close to the characteristic frequencies of the different collective modes of the system.
We derived
in detail the analytic continuation for generic $N$ and discussed, as
an example, the case $N=3$. Power-law singularities are found every
time the frequency is equal to the characteristic frequency of a collective mode ($\omega_j(q) \sim u_j q$), with the same exponent $2\eta-\eta_j-1$, but a different weight  when approaching the singularity from the left or from the right. The power-law singularity replaces the expected delta function for noninteracting particles. Moreover if the characteristic exponent at
the singularity becomes
negative, as in the case of systems with attractive interaction when considering density-density
correlations or in
systems with repulsion when considering bosonic Green's functions, a cusp is expected to replace the power-law divergence.
All these results are valid in the ground state.
For nonzero  temperature $T$, the power-law divergence at $\omega=u_j q$ is replaced by a
maximum diverging as $T^{2\eta-\eta_j-1}$ as $T \to 0$.
In the vicinity
of the maximum, from simple scaling, we expect $\mathrm{Im} \chi(q,\omega) \sim
(k_B T)^{2\eta-\eta_j-1} f_{\text{temp.}} [(\omega -u_j q)/k_B T]$, with
$f_{\text{temp.}} (x\gg 1)\sim x^{2\eta-\eta_j-1}$.
If we now turn to
a zero temperature system
of finite length $L$, we expect the power-law divergence at  $\omega=u_j
q$ to be replaced by a maximum diverging
 as $(1/L)^{2\eta-\eta_j-1}$,
and in the vicinity of the maximum  $\mathrm{Im} \chi(q,\omega) \sim
(1/L)^{2\eta-\eta_j-1} f_{\mathrm{len.}}[L (\omega/u_j- q)]$ with
$f_{\text{len.}}\sim (x\gg 1)\sim x^{2\eta-\eta_j-1}$. If the calculation
of both functions $f_{\mathrm{len.}}$ and $f_{\text{temp.}}$ remains
an open problem, the scaling arguments suggest that the power law
behavior of   $\mathrm{Im} \chi(q,\omega)$ is observable for finite
temperature and finite size provided $|\omega - u_j q| \gg k_B T,
u_j/L$. Such behavior
could be probed by Bragg or time of flight spectroscopy
in the case of atomic gases of mixed species or by
 inelastic neutron scattering technique in one
quantum magnets with orbital and spin modes. An obvious extension of
the results of our manuscript is the calculation of fermion (or anyon)
spectral functions in multicomponent Luttinger liquid. In the case of
a two-component liquid, the fermion spectral functions are expressible
in terms of Appell hypergeometric functions\cite{orignac2011}. For the
case of three or more components, the results of our manuscript hint
that the fermion spectral functions should be expressible as
Srivastava-Daoust hypergeometric functions or a suitable
generalization.

\begin{acknowledgments}
 E. O. thanks the University of Salerno for support during his stay at
 the Department of Physics ``E.R. Caianiello''.
\end{acknowledgments}

\appendix
\section{Analytic continuation of the Appell $F_1$ function}
\label{app:appell-contin}

In the calculation of the response function, an analytic continuation
of the Lauricella hypergeometric function $F_D^{(N-1)}$ is necessary.
In the present
appendix, we present the analytic continuation of the Appell $F_1$
hypergeometric function that corresponds to the particular case of
$N=3$. The analysis of that case is the stepping stone for the case of
general $N$.

In order to find the analytic continuation, we start from the integral
representation of the Appell hypergeometric
function\cite{erdelyi_functions_1,dlmf_appell}
\begin{eqnarray}
  \label{eq:appell-f1-integral}
  F_1(a;b_1,b_2;c;z_1,z_2)=\frac{\Gamma(c)}{\Gamma(a) \Gamma(c-a)}
  \int_0^1 dt \frac{t^{a-1} (1-t)^{c-a-1}}{(1-t z_1)^{b_1} (1-t
    z_2)^{b_2}}
\end{eqnarray}
when $|z_1|,|z_2|<1$, expanding in series
(\ref{eq:appell-f1-integral}) and integrating w.r. t. $t$ gives back
the series expansion. We wish to use (\ref{eq:appell-f1-integral}) to
express $\lim_{\epsilon_1,\epsilon_2 \to 0_+}
F_1(a;b_1,b_2;c;x_1+i\epsilon_1,x_2+i\epsilon_2)$ for $x_1$ and $x_2$
real for the cases $x_2<1<x_1$ and $1<x_2<x_1$.

\subsection{case $x_2<1<x_1$}

In order to calculate the integral (\ref{eq:appell-f1-integral}), we
need to take into account the sole branch cut of $(1-t (x_1+i
\epsilon_1))^{b_1}$ for $t>1/x_1$. Because of this cut, we have for
$t>x_1$, $(1-t (x_1+i \epsilon_1))^{b_1}=e^{- i\pi b_1} (t x_1
-1)^{b_1}$. Therefore, we split the $t$ integral in
(\ref{eq:appell-f1-integral}) into two integrations on $[0,1/x_1]$ and
$[1/x_1,1]$. After a change of variable $t=s/x_1$, we find for the
$[0,x_1]$ integral :
\begin{eqnarray}
  \label{eq:0-to-inv-x1}
  \int_0^{1/x_1} dt \frac{t^{a-1} (1-t)^{c-a-1}}{(1-t x_1)^{b_1} (1-t
    x_2)^{b_2}}  = \frac 1 {x_1^a}
    \frac{\Gamma(a)\Gamma(1-b_1)}{\Gamma(1+a-b_1)}
    F_1(a;a+1-c,b_2;a+1-b_1;1/x_1;x_2/x_1),
\end{eqnarray}
a purely real expression. Then, for the $[1/x_1,1]$ integration, we
find:

\begin{eqnarray}
  \label{eq:inv-x1-to-1}
 \int_{1/x_1}^1 dt \frac{t^{a-1} (1-t)^{c-a-1}}{(1-t (x_1+i 0_+))^{b_1} (1-t
    x_2)^{b_2}}&=& e^{i \pi b_1} \frac{(x_1-1)^{c-a-b_1}}{x_1^{c-b_2-1}
      (x_1-x_2)^{b_2}} \frac{\Gamma(c-a)
      \Gamma(1-b_1)}{\Gamma(1+c-a-b_1)} \nonumber \\ &&  \times
    F_1\left(1-b_1;1-a;b_2;c-a+1-b_1;1-x_1,\frac{x_2(x_1-1)}{x_1-x_2}\right),
\end{eqnarray}
where we have used the linear change of variables
$t=1/x_1+s(1-1/x_1)$.
Our result for the analytic continuation is then:
\begin{eqnarray}
  \label{eq:x2_less_1}
 &&  F_1(a;b_1,b_2;c;x_1+i0_+,x_2)=\frac{\Gamma(c)
    \Gamma(1-b_1)}{\Gamma(c-a) \Gamma(1+a-b_1)} \frac 1 {x_1^a}
  F_1(a;a+1-c,b_2;a+1-b_1;1/x_1;x_2/x_1) \nonumber \\ && + e^{i \pi b_1} \frac{\Gamma(c)
    \Gamma(1-b_1)}{\Gamma(a) \Gamma(1+c-a-b_1)}   \frac{(x_1-1)^{c-a-b_1}}{x_1^{c-b_2-1}
      (x_1-x_2)^{b_2}}
    F_1\left(1-b_1;1-a;b_2;c-a+1-b_1;1-x_1,\frac{x_2(x_1-1)}{x_1-x_2}\right).
\end{eqnarray}

\subsection{case $1<x_2<x_1$}
In that case, we need to consider both the branch cut of $(1-t (x_1+
i0_+))^{b_1}$ and of  $(1-t (x_2+i0_+))^{b_2}$. Thus, we are lead to
split the integral (\ref{eq:appell-f1-integral}) into three
integrations over $[0,1/x_1]$, $[1/x_1,1/x_2]$ and $[1/x_2,1]$. The
first of these integrals is still given by (\ref{eq:0-to-inv-x1}). The
second integral is given by:
\begin{eqnarray}
  \label{eq:inv-x1-to-inv-x2}
  \int_{1/x_1}^{1/x_2} dt \frac{t^{a-1} (1-t)^{c-a-1}}{(1-t (x_1+i 0_+))^{b_1} (1-t
    x_2)^{b_2}} &=& e^{i\pi b_1} \frac{\Gamma(1-b_1)
    \Gamma(1-b_2)}{\Gamma(2-b_1-b_2)} \frac{(x_1-x_2)^{1-b_1-b_2}
    (x_1-1)^{c-a-1}}{x_1^{c-b_2-1} x_2^{1-b_1}} \nonumber \\
  && \times
  F_1\left(1-b_1;1-a,1+a-c;2-b_1-b_2;1-\frac{x_1}{x_2},
    \frac{x_1-x_2}{x_2(x_1-1)}\right),
\end{eqnarray}
where we have used the change of variable $t=1/x_1+s(1/x_2-1/x_1)$ and
taken into account the branch cut of $(1-t (x_1 +i 0_+))^{b_1}$.
For the third integral, we have:
\begin{eqnarray}
  \label{eq:inv-x2-to-1}
  \int_{1/x_2}^1 dt \frac{t^{a-1} (1-t)^{c-a-1}}{(1-t (x_1+i 0_+))^{b_1} (1-t
    (x_2+i0_+) )^{b_2}} &=& e^{i\pi (b_1+b_2)} \frac{\Gamma(1-b_2)
    \Gamma(c-a)}{\Gamma(1+c-a-b_2)}
  \frac{(x_2-1)^{c-a-b_2}}{x_2^{a-b_1} (x_1-x_2)^{b_1}} \nonumber \\
  && \times
  F_1\left(1-b_2;1-a,b_1;1+c-a-b_2;1-x_2,\frac{x_1(1-x_2)}{x_1-x_2}\right),
\end{eqnarray}
where we have taken both branch cuts into account, and we have used
the change of variables $t=1/x_2+(1-1/x_2)s$. The final result is:
\begin{eqnarray}
  \label{eq:x_1-x_2-gt1}
&&  F_1(a,b_1,b_2;c;x_1+i0_+,x_2+i0_+)=\frac{\Gamma(c)\Gamma(1-b_1)}{\Gamma(c-a)
    \Gamma(1+a-b_1)} \frac 1 {x_1^a}
  F_1\left(a;1+a-c,b_2;1+a-b_1;\frac 1 {x_1},\frac{x_2}{x_1}\right)
  \nonumber \\
&&+ e^{i\pi b_1}\frac{\Gamma(c)\Gamma(1-b_1)\Gamma(1-b_2)}{\Gamma(a)
  \Gamma(c-a) \Gamma(2-b_1-b_2)} \frac{(x_1-x_2)^{1-b_1-b_2}
  (x_1-1)^{c-a-1}}{x_1^{c-b_2-1} x_2^{1-b_1}}
F_1\left(1-b_1;1-a,1+a-c;2-b_1-b_2;1-\frac{x_1}{x_2},\frac{x_1-x_2}{x_2(x_1-1)}\right)
\nonumber \\
&& + e^{i\pi(b_1+b_2)} \frac{\Gamma(c) \Gamma(1-b_2)} {\Gamma(a)
  \Gamma(1+c-a-b_2)} \frac{(x_2-1)^{c-a-b_2}}{x_2^{a-b_1}
  (x_1-x_2)^{b_1}}
F_1\left(1-b_2;1-a,b_1;1+c-a-b_2;1-x_2,\frac{x_1(1-x_2)}{x_1-x_2}\right)
\end{eqnarray}

For the case $N=3$, the expressions (\ref{eq:x2_less_1}) and
(\ref{eq:x_1-x_2-gt1}) must be injected in the
integral~(\ref{eq:feynnman-final}) after analytic continuation
to yield the response function.

\section{Analytic continuation of the Lauricella $F_D$ function}
\label{app:lauricella-fd-contin}

The Lauricella $F_D$ function has the integral representation:
\begin{eqnarray}
  \label{eq:lauricella-fd-integral}
  F_D(a;b_1,\ldots,b_n;c;z_1,\ldots,z_n)=\frac{\Gamma(c)}{\Gamma(a)
    \Gamma(c-a)} \int_0^1 dt \frac{t^{a-1}
    (1-t)^{c-a-1}}{\prod_{j=1}^n (1-z_j t)^{b_j}}
\end{eqnarray}

If we want to calculate for $x_n<x_{n-1}<\ldots<x_2<x_1$:
\begin{eqnarray}
  \label{eq:on-cut-fd}
  \lim_{\epsilon_j \to 0_+}
  F_D(a;b_1,\ldots,b_n;c;x_1+i\epsilon_1,\ldots,x_n+i\epsilon_n)
\end{eqnarray}
we have to consider the cuts of the functions $(1-t x_j +i
0_-)^{-b_j}$. We are thus led to consider separately $n$ cases, i. e.
$x_n<\ldots<x_{j+1}<1<x_{j}<\ldots<x_1$ with $j=1,\ldots,n-1$ and
$1<x_n<\ldots<x_1$.

In the case of $x_n<\ldots<x_{j+1}<1<x_{j}<\ldots<x_1$, the integral
(\ref{eq:lauricella-fd-integral}) has
to be split into $j+1$ integrations over the intervals
$[0,1/x_1]$,$[1/x_1,1/x_2]$,\ldots,$[1/x_{j},1]$. In analogy to the
case of the Appell $F_1$ function, each
integral gives a contribution proportional to a Lauricella $F_D$
function. The resulting expression is:
\begin{eqnarray}
  \label{eq:lauricella_continuation}
&& F_D^{(N)}(a;b_1,\ldots,b_N;c;x_1+i0,\ldots,x_N+i0) = \nonumber \\
&& = \frac{\Gamma(c)}{\Gamma(a) \Gamma(c-a)} \left[ \frac{\Gamma(a)
    \Gamma(1-b_1)}{\Gamma(1+a-b_1) x_1^a}
  F_D^{(N)}\left(a;1+a-c,b_2,\ldots,b_N;1+a-b_1;\frac 1 {x_1},\frac
    {x_2}{x_1}\ldots,\frac{x_N}{x_1}\right) \right. \nonumber \\
&&  + \sum_{m=1}^{j-1} \frac{\Gamma(1-b_m)
  \Gamma(1-b_{m+1})}{\Gamma(2-b_m-b_{m+1})}
\frac{(x_m-x_{m+1})^{1-b_m-b_{m+1}} (1-x_m)^{c-a-1}}{x_m^{c-1-b_{m+1}}
  x_{m+1}^{1-b_m}} \prod_{l\ne m,m+1}
\left|\frac{x_l-x_m}{x_m}\right|^{-b_l} e^{i \pi \sum_{l=1}^{m} b_l}
\times \nonumber \\
&&   F_D^{(N)}\left(1-b_m;b_1,\ldots,b_{m-1},1-a,1+a-c,b_{m+2},\ldots,b_N;2-b_{m}-b_{m+1};\frac{1-\frac{x_m}{x_{m+1}}
     }{1-\frac{x_m}{x_1}},\ldots,\frac{1-\frac{x_m}{x_{m+1}}
     }{1-\frac{x_m}{x_{m-1}}},1-\frac{x_m}{x_{m+1}},\frac{1-\frac{x_m}{x_{m+1}}
     }{1-x_m},\right. \nonumber \\ && \left. \frac{1-\frac{x_m}{x_{m+1}}
     }{1-\frac{x_m}{x_{m+2}}},\ldots,\frac{1-\frac{x_m}{x_{m+1}}
     }{1-\frac{x_m}{x_{N}}}\right) \nonumber \\
&&
+ \frac{\Gamma(1-b_j) \Gamma(c-a)}{\Gamma(1+c-a-b_j)}
\frac{(x_j-1)^{c-a-b_j}}{x_j^{c-1}}  \prod_{l\ne j}
\left|\frac{x_l-x_j}{x_j}\right|^{-b_l}  e^{i \pi \sum_{l=1}^{j}
  b_l}\times \nonumber \\
&& \left.  F_D^{(N)}\left(1-b_j;b_1,\ldots,b_{j-1},1-a,b_{j+1},\ldots,b_N;1+c-a-b_{j};\frac{1-x_j}{1-\frac{x_j}{x_1}},
    \ldots,\frac{1-x_j}{1-\frac{x_j}{x_{j-1}}},1-x_j,\frac{1-x_j
     }{1-\frac{x_j}{x_{j+1}}},\ldots,\frac{1-x_j}
     {1-\frac{x_j}{x_{N}}}\right) \right]
\end{eqnarray}
By reducing to $N=2$ it can be checked that the results of
Sec.~\ref{app:appell-contin} are recovered.

 In the case $1<x_1<\ldots<x_n$, we have to split the integral
(\ref{eq:lauricella-fd-integral}) into $n+1$ integrations over the
intervals
$[0,1/x_1]$,\ldots,$[1/x_j,1/x_{j+1}]$,\ldots,$[1/x_n,1]$. Each
integration contributes a term proportional to a Lauricella $F_D$ function.

\section{Asymptotic expansion of the Appell $F_1$ function}
\label{app:asymp-f1}

We wish to obtain an asymptotic expansion of the Appell function
$F_1(a;b_1,b_2;c;x_1,x_2)$ in the case of $x_2<0$ and $x_1 \to 1_-$.
First, we need to obtain an expression of the Appell function in the
form of a convergent series for all $x_2<0$.
We consider the series expansion for the Appell hypergeometric
function $F_1$:
\begin{eqnarray}
  \label{eq:f1-series}
  F_1(a;b_1,b_2;c;x_1,x_2)&=&\sum_{n_1,n_2} \frac{(a)_{n_1+n_2}
    (b_1)_{n_1} (b_2)_{n_2}}{(c)_{n_1+n_2}} \frac{x_1^{n_1}}{n_1 ! }
  \frac{x_2^{n_2}}{n_2 !}, \\
 &=& \sum_{n_1}  \frac{(a)_{n_1} (b_1)_{n_1}}{(c)_{n_1}}
 \frac{x_1^{n_1}}{n_1 ! } {}_2F_1(a+n_1,b_2;c+n_1;x_2),
\end{eqnarray}
where we have used the notation\cite{abramowitz_math_functions}:
\begin{eqnarray}
  (a)_n=\frac{\Gamma(n+a)}{\Gamma(a)}
\end{eqnarray}
Using the second line of (\ref{eq:f1-series}), we can define the
function $F_1$ for all $x_2 \notin [1,+\infty[$. For $x_2<0$, we can
use Eq. (15.3.4) from Ref.~\onlinecite{abramowitz_math_functions}, to
rewrite:
\begin{eqnarray}
  F_1(a;b_1,b_2;c;x_1,x_2)&=& (1-x_2)^{-b_2} \sum_{n_1}  \frac{(a)_{n_1} (b_1)_{n_1}}{(c)_{n_1}}
 \frac{x_1^{n_1}}{n_1 ! }
 {}_2F_1\left(b_2,c-a;c+n_1;\frac{x_2}{x_2-1}\right), \nonumber \\
&=& (1-x_2)^{-b_2}  \sum_{n_1,n_2} \frac{(a)_{n_1}
    (b_1)_{n_1} (b_2)_{n_2} (c-a)_{n_2}}{(c)_{n_1+n_2}} \frac{x_1^{n_1}}{n_1 ! }
  \frac{1}{n_2 !} \left(\frac{x_2}{x_2-1}\right)^{n_2}, \nonumber \\
&=& (1-x_2)^{-b_2}
F_3\left(a,c-a;b_1,b_2;c;x_1,\frac{x_2}{x_2-1}\right),  \\
&=&  (1-x_2)^{-b_2} \sum_{n_2} \frac{(c-a)_{n_2}
  (b_2)_{n_2}}{(c)_{n_2} n_2 !}  \left(\frac{x_2}{x_2-1}\right)^{n_2}
{}_2F_1(a,b_1;c+m;x_1)
\end{eqnarray}
where we have used the convergence of the Gauss hypergeometric series
to obtain the last two lines. with this
convergent series, we can analyze its behavior as $x_1 \to 1_-$.
First, when $c+m-a-b_1>0$, we have from Eq.~(15.1.20) in
Ref.\onlinecite{abramowitz_math_functions}:
\begin{eqnarray}
  \lim_{x\to 1_-} {}_2F_1(a,b_1;c+m;x_1) = \frac{\Gamma(c+m)
    \Gamma(c+m-a-b_1)}{\Gamma(c+m-a) \Gamma(c+m-b_1)}.
\end{eqnarray}
When $c+m-a-b_1<0$, using first Eq.~(15.3.3)  in
Ref.\onlinecite{abramowitz_math_functions}, we find that as $x\to
1_-$,
\begin{equation}
   {}_2F_1(a,b_1;c+m;x_1) \sim (1-x_1)^{c-a-b_1+m} \frac{\Gamma(c+m)
       \Gamma(a+b_1-c-m)}{\Gamma(a) \Gamma(b_1)}
\end{equation}
Thus, when $c-a-b_1<0$, we have that:
\begin{eqnarray}
\label{eq:asymp-f1-power}
  F_1(a;b_1;b_2;c;x_1,x_2) \sim (1-x_2)^{-b_2} \frac{\Gamma(c)
       \Gamma(a+b_1-c)}{\Gamma(a) \Gamma(b_1)}   (1-x_1)^{c-a-b_1},
\end{eqnarray}
while for $c-a-b_1>0$,
\begin{eqnarray}
\label{eq:asymp-f1-finite}
  \lim_{x_1\to 1_-}  F_1(a;b_1;b_2;c;x_1,x_2) = (1-x_2)^{-b_2}  \frac{\Gamma(c)
    \Gamma(c+a-b_1)}{\Gamma(c-a) \Gamma(c-b_1)} {}_2 F_1
  \left(c-a-b_1,b_2;c-b_1; \frac{x_2}{x_2 -1}\right)
\end{eqnarray}

Let us now return to the case of a general $N$.
We have to consider the integral:
\begin{eqnarray}
  I^{(N-1)}(q,\omega)=\int_0^1 dt t^{-\eta} (1-t)^{\eta-3/2} F_D^{(N-1)}\left(\frac 1 2;\{\eta_j\}_{1\le j \le
      N-1};\eta;\left\{\frac{u_N^2-u_j^2}{u_N^2} \left(1+\frac{t}{\frac{u_N^2 q^2}{(\omega+i0)^2}
        -1}\right) \right\}_{1\le j \le
      N-1}\right)
\end{eqnarray}

\section{Analytic continuation of the Matsubara correlator for generic
  $\eta$ and N components}\label{app:general-n}
 With the help of
Eq.~(\ref{eq:lauricella_continuation}) we can write for
$t_l<t<t_{l+1}$:
\begin{eqnarray}
  \label{eq:fd-as-sum}
 && F_D^{(N-1)}\left(\frac 1 2;\{\eta_j\}_{1\le j \le
      N-1};\eta;\left\{\frac{u_N^2-u_j^2}{u_N^2} \left(1+\frac{t}{\frac{u_N^2 q^2}{(\omega+i0)^2}
        -1}\right) \right\}_{1\le j \le
      N-1}\right) =\sum_{m=0}^{l-1} \varphi_m(t) +\psi_l(t),
\end{eqnarray}
where:
\begin{eqnarray}
  \label{eq:phi-0}
  && \varphi_0(t)=\frac{\Gamma(\eta) \Gamma(1-\eta_1)}{\Gamma(\eta-1/2)
    \Gamma(3/2-\eta_1)}
  \frac{1}{\left[\left(1-\frac{u_1^2}{u_N^2}\right)\left(1+\frac{t}{\frac{u_N^2 q^2}{(\omega+i0)^2}
        -1}\right)\right]^{1/2}} \times \nonumber \\ &&   F_D^{(N-1)}\left(\frac 1 2;\frac 3 2
  -\eta,\{\eta_j\}_{2\le j \le
      N-1};\frac 3 2 -\eta_1; \frac{1}{\left(1-\frac{u_1^2}{u_N^2}\right)\left(1+\frac{t}{\frac{u_N^2 q^2}{\omega^2}
        -1}\right)}, \left\{\frac{u_N^2-u_j^2}{u_N^2-u_1^2} \right\}_{2\le j \le
      N-1}\right),
\end{eqnarray}
for $m\ge 1$:
\begin{eqnarray}
  \label{eq:phi-m}
 && \varphi_m(t)= \frac{\Gamma(\eta) \Gamma(1-\eta_m)
    \Gamma(1-\eta_{m+1})}{\Gamma(1/2) \Gamma(\eta-1/2)
    \Gamma(2-\eta_m--\eta_{m+1})}
  \left(\frac{u_{m+1}^2-u_m^2}{u_N^2-u_{m+1}^2}\right)^{1-\eta_m}
 \left(\frac{u_{m+1}^2-u_m^2}{u_N^2-u_{m}^2}\right)^{-\eta_{m+1}}
\prod_{k\ne m,m+1} \left|
  \frac{u_{m}^2-u_k^2}{u_N^2-u_{m}^2}\right|^{-\eta_k} e^{i\pi
  \sum_1^m \eta_k} \nonumber \\
 && \frac   {\left[1-\left(1-\frac{u_m^2}{u_N^2}\right)\left(1+\frac{t}{\frac{u_N^2 q^2}{\omega^2}
        -1}\right)\right]^{\eta-3/2}} {\left[\left(1-\frac{u_m^2}{u_N^2}\right)\left(1+\frac{t}{\frac{u_N^2 q^2}{\omega^2}
        -1}\right)\right]^{\eta-1}}
F_D^{(N-1)}\left(1-\eta_m,\{\eta_k\}_{1\le k \le m-1},\frac 1 2,\frac 1 2
-\eta, \{\eta_k\}_{m+2\le k \le
  N-1};2-\eta_m-\eta_{m+1};\right. \\
&& \left. \left\{\frac{(u_m^2-u_{m+1}^2)(u_N^2-u_k^2)}{(u_m^2-u_k^2)(u_N^2-u_{m+1}^2)}\right\}_{1\le
  k \le m-1},    \frac{(u_m^2-u_{m+1}^2)}{(u_N^2-u_{m+1}^2)},\frac{\frac{(u_{m+1}^2-u_{m}^2)}{(u_N^2-u_{m+1}^2)}}{\left(1-\frac{u_m^2}{u_N^2}\right)\left(1+\frac{t}{\frac{u_N^2 q^2}{\omega^2}
        -1}\right)}, \left\{\frac{(u_m^2-u_{m+1}^2)(u_N^2-u_k^2)}{(u_m^2-u_k^2)(u_N^2-u_{m+1}^2)}\right\}_{m+2\le
  k \le N-1}\right)   \nonumber
\end{eqnarray}
and:
\begin{eqnarray}
  \label{eq:psi-l}
&&  \psi_l(t)=\frac{\Gamma(1-\eta_l)\Gamma(\eta-1/2)}{\Gamma(\eta-\eta_l+1/2)}
  \prod_{k \ne l}
  \left|\frac{u_l^2-u_k^2}{u_N^2-u_k^2}\right|^{-\eta_k}
  e^{i\pi\sum_1^l \eta_k}
  \frac {\left[\left(1-\frac{u_l^2}{u_N^2}\right)\left(1+\frac{t}{\frac{u_N^2 q^2}{\omega^2}
        -1}\right)-1\right]^{\eta-\eta_l-1/2}} {\left[\left(1-\frac{u_m^2}{u_N^2}\right)\left(1+\frac{t}{\frac{u_N^2 q^2}{\omega^2}
        -1}\right)\right]^{\eta-1}} \times \\ &&  F_D^{(N-1)}\left(1-\eta_l,\{\eta_k\}_{1\le k \le l-1},\frac 1 2, \{\eta_k\}_{l+1\le k \le
  N-1};\frac 1 2 +\eta-\eta_{l}; \left\{\frac{u_N^2-u_k^2}{u_N^2-u_l^2}\left(1-\left(1-\frac{u_l^2}{u_N^2}\right)\left(1+\frac{t}{\frac{u_N^2 q^2}{\omega^2}
        -1}\right)\right)\right\}_{1\le k\le N-1}\right)    \nonumber
\end{eqnarray}

So that for $u_j q <\omega <u_{j+1} q$:
\begin{eqnarray}
 && I^{(N-1)}(q,\omega)=\sum_{l=1}^{j-1} \int_{t_l}^{t_{l+1}} dt
  t^{-\eta}(1-t)^{\eta-3/2} \left(\sum_{m=0}^{l-1} \varphi_m(t) +
    \psi_l(t)\right) + \int_{t_j}^1 dt
  t^{-\eta}(1-t)^{\eta-3/2} \left(\sum_{m=0}^{j-1} \varphi_m(t) +
    \psi_j(t)\right) \nonumber \\
&& + \int_0^{t_1} dt t^{-\eta} (1-t)^{\eta-3/2} F_D^{(N-1)}\left(\frac 1 2;\{\eta_j\}_{1\le j \le
      N-1};\eta;\left\{\frac{u_N^2-u_j^2}{u_N^2} \left(1+\frac{t}{\frac{u_N^2 q^2}{\omega^2}
        -1}\right) \right\}_{1\le j \le
      N-1}\right)
\end{eqnarray}
We can rearrange that sum into:
\begin{eqnarray}
\label{eq:int_gen_N}
&&  I^{(N-1)}(q,\omega)=\sum_{m=0}^{j-1} \int_{t_{m+1}}^1 dt
 t^{-\eta}(1-t)^{\eta-3/2} \varphi_m(t) +  \sum_{l=1}^{j-1} \int_{t_l}^{t_{l+1}} dt
  t^{-\eta}(1-t)^{\eta-3/2}  \psi_l(t) + \int_{t_j}^{1} dt
  t^{-\eta}(1-t)^{\eta-3/2}  \psi_j(t)  \nonumber \\
&& + \int_0^{t_1} dt t^{-\eta} (1-t)^{\eta-3/2} F_D^{(N-1)}\left(\frac 1 2;\{\eta_j\}_{1\le j \le
      N-1};\eta;\left\{\frac{u_N^2-u_j^2}{u_N^2} \left(1+\frac{t}{\frac{u_N^2 q^2}{\omega^2}
        -1}\right) \right\}_{1\le j \le
      N-1}\right)
\end{eqnarray}
So we have to calculate $2j$ integrals. When in the last integral we do substitute the expression of  (\ref{eq:psi-l}), a factor $\left( \omega^2-(u_jq)^2\right)^{2\eta-\eta_j-1}$ appears with an integral which is regular in the limit $\omega \rightarrow u_jq$, so the asymptotic behavior previously predicted is recovered. The first class of  integrals in (\ref{eq:int_gen_N}) can be, instead, manipulated  by using the definition (\ref{eq:phi-m}) and performing the following change of variables:
\begin{equation}
t=\frac{\left(\frac{(u_N q)^2}{\omega^2}-1\right)}{\left(\frac{(u_N )^2}{\omega_{m+1}^2}-1\right)}+\left(1-\frac{\left(\frac{(u_N q)^2}{\omega^2}-1\right)}{\left(\frac{(u_N )^2}{\omega_{m+1}^2}-1\right)}\right)s.
\end{equation}
The integrals turn out to be regular for $\omega>u_j q$ and for
$\omega \rightarrow u_jq+0$ they behave as  $\left(
  \omega^2-(u_jq)^2\right)^{\eta-1/2}$ giving only a subdominant
contribution.

\end{document}